\documentclass[preprint,prd,showpacs]{revtex4} 

\usepackage{graphicx}
 
\begin{document}

\title{Negative Energy Seen By Accelerated Observers}

\author{L. H. Ford}
\email{ford@cosmos.phy.tufts.edu}
\affiliation{Institute of Cosmology, Department of Physics and
  Astronomy, 
Tufts University, Medford, Massachusetts 02155, USA}
\author{Thomas A. Roman}
\email{roman@ccsu.edu}
\affiliation{
Department of Mathematical Sciences, Central Connecticut State
University, 
New Britain, Connecticut 06050, USA}

\begin{abstract}
The sampled negative energy density seen by inertial observers, 
in arbitrary quantum states is limited by  quantum inequalities,
which take the form of an inverse relation 
between the magnitude and duration of the negative energy. 
The quantum inequalities severely limit the utilization of negative energy
to produce gross macroscopic effects, such as violations of the second 
law of thermodynamics. The restrictions on the sampled energy density
along the worldlines of accelerated observers are much weaker than
for inertial observers. Here we will illustrate this with several explicit
examples. We consider the worldline of a particle undergoing sinusoidal
motion in space in the presence of a single mode squeezed vacuum
state of the electromagnetic field. We show that it is possible for
the integrated energy density along such a worldline to become
arbitrarily negative at a constant average rate. Thus the
averaged weak energy condition is violated in these examples.
This can be the case even when the particle moves at non-relativistic 
speeds. We use the Raychaudhuri equation to show that there can be 
net defocussing of a congruence of these accelerated worldlines.
This defocussing is an operational signature of the negative
integrated energy density. These results in no way 
invalidate nor undermine either the validity or utility of the 
quantum inequalities for inertial observers. In particular, they do not change 
previous constraints on the production of macroscopic effects with negative 
energy, e.g., the maintenance of traversable wormholes.
 
\end{abstract}


\maketitle

\section{Introduction}
\label{sec:intro}

It is well known that quantum field theory allows for the existence of negative energy density,
which constitute local violations of the weak energy condition. For a recent review, see Ref.~\cite{F10}. 
Negative energy density can arise either from
boundaries, as in the Casimir effect, from background spacetime curvature, or from selected
quantum states in Minkowski spacetime. The last possibility will be the focus of the present
paper.  It is possible to create states, such as a squeezed vacuum state of the quantized electromagnetic
field, in which the energy density at a given spacetime point is arbitrarily negative. However, the
duration of the negative energy is strongly constrained by quantum 
inequalities~\cite{F78,F91,FR95,FR97,Flanagan97,FE98,P02,FH05}. These are restrictions on a time
averaged energy density measured by an observer. (Time averaging is essential, as there is no
analogous restriction on spatial averages~\cite{FHR02}.) Let us consider the case of inertial observers
in Minkowski spacetime, with four velocity $u^\mu$. If $ \langle T_{\mu \nu}  \rangle$ is the
expectation value of the normal ordered stress tensor operator in an arbitrary quantum state,
then quantum inequalities take the form
\begin{equation}
\int_{-\infty}^\infty f(\tau)\, \langle T_{\mu \nu} \rangle\, u^{\mu}u^{\nu}  \, d\tau \geq
-\frac{C_0}{{\tau_0}^d} \,.
 \label{eq:QI}
\end{equation} 
Here $\tau$ is the observer's proper time, $ f(\tau)$ is a sampling function with characteristic width
$\tau_0$, and $d$ is the number of spacetime dimensions. The dimensionless constant $C_0$ depends
upon the form of the sampling function, and is typically small compared to unity. 
In the limit $\tau_0 \rightarrow \infty$, Eq.~(\ref{eq:QI}) becomes the averaged weak energy condition
 \begin{equation}
\int_{-\infty}^\infty  \langle T_{\mu \nu} \rangle\, u^{\mu}u^{\nu}  \, d\tau \geq 0\,,
\end{equation}
which states that the integrated energy density along an inertial worldline is non-negative.
The essence of a quantum inequality is that there is an inverse relation between the magnitude 
and duration of negative energy density. These relations place strong constraints on the effects 
of  negative energy for violating the second law of thermodynamics~\cite{F78}, and for maintaining traversable 
wormholes~\cite{FR96} or warpdrive spacetimes~\cite{PF97}. 

A more general quantum inequality for arbitrary worldlines has been proven by Fewster~\cite{Fewster00}.
However, this inequality is often very difficult to evaluate explicitly and can be very weak. There are 
some known examples where the integrated energy density along a non-inertial world line can be
arbitrarily negative. One example comes from the Fulling-Davies moving mirror model in two spacetime
dimensions~\cite{FD76,DF77}. A mirror with increasing proper acceleration 
to the right can emit a steady
flux of negative energy to the right. An inertial observer could only  see this negative energy for a finite time
before being hit by the mirror, and the integrated energy density seen will be consistent with Eq.~(\ref{eq:QI}).
However, an accelerated observer who stays ahead of the mirror can see an arbitrary amount of negative
energy. This example suffers from two unrealistic features: it can only be formulated in two spacetime
dimensions, and it requires an observer with ever increasing proper acceleration.

A second example was provided by Fewster and Pfenning~\cite{FP06}, who 
analyzed the case of a uniformly accelerating observer in the Rindler 
vacuum state. This state has negative energy everywhere  
within the Rindler wedge. An observer with constant acceleration can also see an arbitrary amount of negative
energy. However, the constant acceleration requires the observer to move arbitrarily close
to the speed of light and hence have an unlimited source of energy. It is also not clear whether
the Rindler vacuum is a physically realizable state.

The main purpose of this paper is to construct some more realistic examples of accelerated motion
in which the observer can have arbitrarily negative integrated energy density. We will consider
observers who undergo sinusoidal motion in the presence of a squeezed vacuum state of the quantum
electromagnetic field. We find that even in the case of non-relativistic motion, it is possible for the
integrated energy density in such an observer's frame to grow negatively at a constant rate in time.
In Sect.~\ref{sec:planewave}, we consider a squeezed vacuum state for a single plane wave mode,
and motions both perpendicular and parallel to the direction of propagation of the wave. In 
Sect.~\ref{sec:cavity}, we repeat the analysis for the lowest mode in a resonant cavity in a squeezed 
vacuum state. In Sect.~\ref{sec:focus}, we address a possible physical effect of accumulating negative
energy density, in the form of defocussing of a congruence of accelerated worldlines. Our results
are summarized and discussed in Sect.~\ref{sec:final}. In particular, we argue that the results in this 
paper neither contradict, nor diminish the utility of, the usual quantum  inequalities proven for 
inertial observers. 

Throughout this paper, units in which $\hbar = c =1$ will be used. Electromagnetic quantities 
are in Lorentz-Heaviside units.

\section{Oscillations Through a Plane Wave}
\label{sec:planewave}

Let us first evaluate the stress tensor components for a single mode plane wave in a squeezed vacuum state 
of the electromagnetic field.  The electromagnetic stress tensor is given in terms of the field strength tensor as
\begin{equation}
T_{\alpha \beta}=F_{\alpha \rho}{F_{\beta}}^ {\rho}
-\frac{1}{4} g_{\alpha \beta} \,F_{\mu \nu}F^{\mu \nu} \,.
\label{eq:T}
\end{equation}
Its spatial components are 
\begin{equation}
T_{j l} = -E_j E_l - B_j B_l + \frac{1}{2} \delta_{jl} ( {\mathbf E}^2 +  {\mathbf B}^2) \,, 
\end{equation} 
the energy density is
\begin{equation}
T^{tt } = \frac{1}{2}  ( {\mathbf E}^2 +  {\mathbf B}^2) \,,
\end{equation}
and the energy flux in the $i$-direction is 
\begin{equation}
T^{ti}=  (\mathbf{E \times B})^i 
\end{equation}
Write the electric and magnetic field operators in terms of photon creation operators
${ {\hat a}^\dagger}_{{\bf k} \lambda}$ and annihilation operators ${\hat a}_{{\bf k} \lambda}$   as
\begin{equation}
{\mathbf E} = {\sum_{{\bf k},\lambda}}\, ( {\hat a}_{{\bf k}\lambda} 
\,   {\bf \mathcal E}_{{\bf k} \lambda} +
{ {\hat a}^\dagger}_{{\bf k} \lambda} \,  
{{\bf \mathcal E}^\ast}_{{\bf  k} \lambda}) \,,
\label{eq:E-op}
\end{equation}
and
\begin{equation}
{\mathbf B} = {\sum_{{\bf k},\lambda}}\, ( {\hat a}_{{\bf k}\lambda} 
\,   {\bf \mathcal B}_{{\bf k} \lambda} +
{ {\hat a}^\dagger}_{{\bf k} \lambda} \,  
{{\bf \mathcal B}^\ast}_{{\bf  k} \lambda}) \,.
\label{eq:B-op}
\end{equation}

Assume that the excited mode is a plane wave propagating in the $z$-direction, with 
polarization in the $x$-direction. Then its mode functions take the form
$ {\bf \mathcal E}_{{\bf k} \lambda} = \hat{\bf x} \,  { \mathcal E}$, and
$ {\bf \mathcal B}_{{\bf k} \lambda} = \hat{\bf y} \,  { \mathcal B}$, where 
\begin{equation}
{\mathcal E} = {\mathcal B } = {\sqrt{\frac{\Omega}{2 V}}}  \, e^{i \Omega (z-t)} \,.
\label{eq:mode}
\end{equation}
Here $V$ is the quantization volume and $\Omega = |{\bf k}|$ 
is the angular frequency of the wave. 
Quadratic operators are assumed to be normal ordered with respect to the Minkowski
vacuum state, so 
\begin{equation}
\langle E_j E_l \rangle = \delta_{jx}\, \delta_{lx} \,  \langle  {\mathbf E}^2 \rangle =
{\mathcal E}^2 \langle {\hat a}^2 \rangle +
({\mathcal E^\ast})^2  \langle {( {\hat a}^\dagger})^2 \rangle + 
2 |{\mathcal E}|^2 \langle   {\hat a}^\dagger  {\hat a}  \rangle \,.
\end{equation}
where $ {\hat a}$ is the annihilation operator for the excited mode. Similarly,
\begin{equation}
\langle B_j B_l \rangle = \delta_{jy}\, \delta_{ly} \,  \langle  {\mathbf E}^2 \rangle \,.
\end{equation}
The quantum state is taken to be a single mode in which case
\begin{eqnarray}
\langle  {\mathbf E}^2 \rangle =  \langle  {\mathbf B}^2 \rangle & =& 
 2 \, {\mathcal Re}\, [ {{\rm sinh}^2}r \,   |{\mathcal E}|^2   -   
{\mathcal E}^2 \, {\rm sinh}r \,  {\rm cosh}r \,  e^{i \delta}) ]   \nonumber \\ 
&=& \frac{\Omega}{V} \, {\rm sinh}r \,\biggl \{  {{\rm sinh}}r  -  
 {\rm cosh}r \, {\rm cos}[2 \Omega (z-t) + \delta] \biggr \}  \,,
\label{eq:Exsq}
\end{eqnarray}
where $r$ is the ``squeeze parameter'' and $\delta$ is a phase parameter. 
The nonzero components of the stress tensor are given by
\begin{equation}
\langle T^{tt} \rangle = \langle T^{zz} \rangle = \langle T^{tz} \rangle 
= \langle  {\mathbf E}^2 \rangle\,.
\label{eq:Tcomp}
\end{equation}
We see from 
Eqs.~(\ref{eq:Exsq}) and (\ref{eq:Tcomp}) that the energy density can be periodically negative in the 
lab (i.e., inertial) observer's frame, but the positive energy density always outweighs the negative energy density, 
in accordance with the quantum inequalities.

The energy density in the inertial frame has its minimum (most negative) value when the cosine term in
Eq.~(\ref{eq:Exsq}) is one, so
\begin{equation}
\langle T^{tt} \rangle \geq  - \frac{\Omega}{V} \, \sinh r \, (\cosh r - \sinh r) = - \frac{\Omega}{2V} \, 
(1- {\rm e}^{-2r})   > - \frac{\Omega}{2V} \, .
\end{equation} 
Thus the maximally negative energy density is bounded below, and occurs for large $r$. However, in
this limit the maximally positive energy density is unbounded and grows as $ {\rm e}^{2r}$. In the opposite limit,
where $r \ll 1$, the energy density is approximately oscillatory 
\begin{equation}
\langle T^{tt} \rangle \approx -  r\,\frac{\Omega}{V} \,  {\rm cos}[2 \Omega (z-t) + \delta]  
+r^2\,\frac{\Omega}{V}  + O(r^3) \,.
\label{eq:small_r}
\end{equation}
However, there is also a positive non-oscillatory term of order $r^2$.

\subsection{Perpendicular Motion}
\label{sec:perpmotion}

Now consider a non-geodesic observer who moves on a path 
which is perpendicular to the direction of propagation of the wave. 
Let this path be defined by   
\begin{equation}
v^x(t) = \frac{dx}{dt} = A \,{\rm sin}(\omega t) \,,
\end{equation}
where $|A|<1$, and $v^y=v^z=0$, and $\omega$ is the  angular oscillation frequency of the 
observer's motion, and where we have chosen $z=0$. Then
\begin{equation}
\gamma = \frac{1}{\sqrt{1-v^2}} =  \frac{1}{\sqrt{1-A^2 {\rm sin}^2(\omega t)}} \,,
\label{eq:gamma}
\end{equation}
and the observer's four-velocity (as measured in the lab frame) is
\begin{equation}
u^{\mu}=\gamma (1,v^x,0,0) \,,
\end{equation}
where $u^t = \gamma ={dt}/{d\tau}$. 

\begin{figure} 
\begin{center} 
\includegraphics[width=5in]{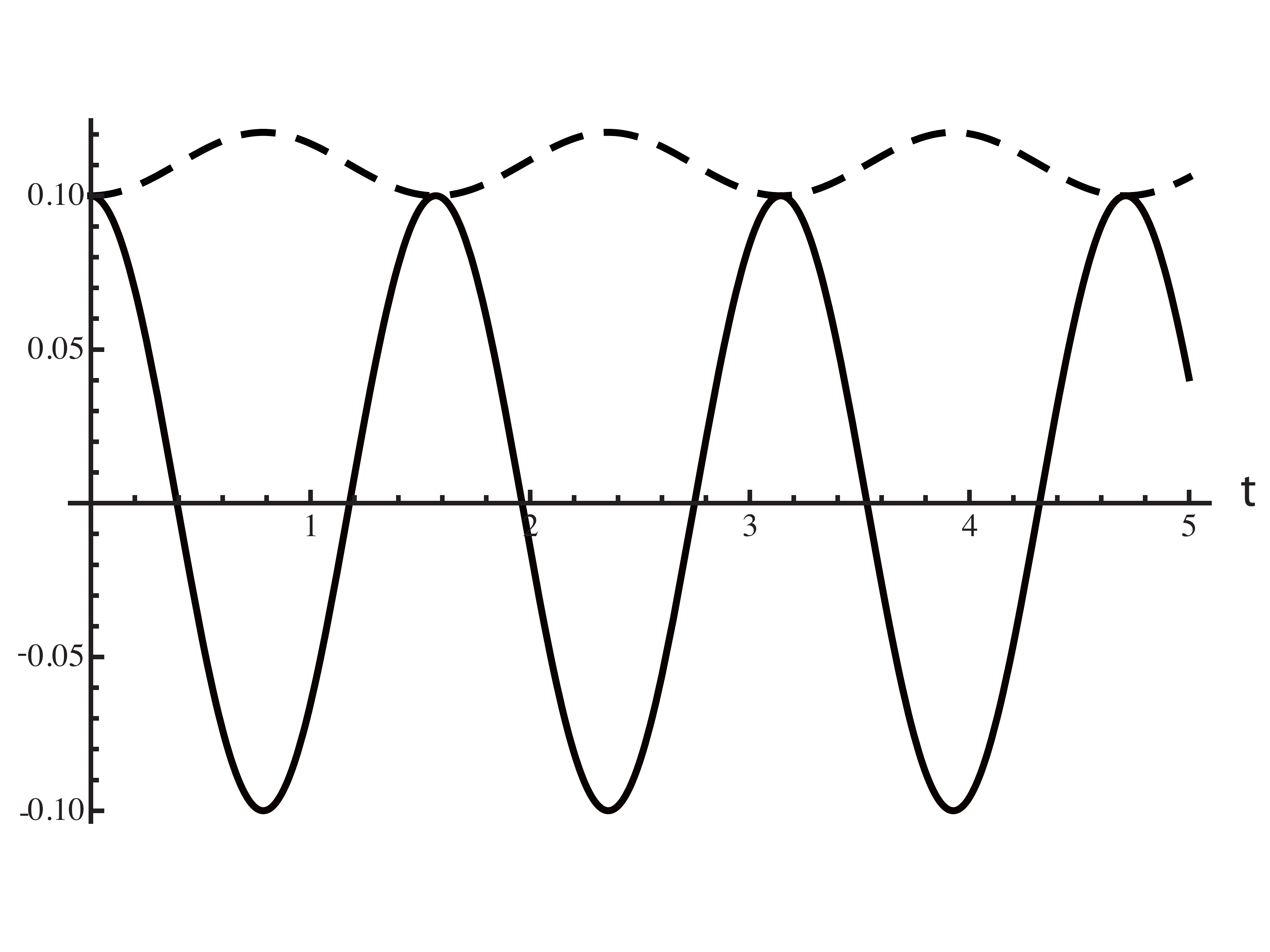}
\end{center} 
\vspace{-0.5in}
\caption{The figure illustrates that 
maximum negative energy density is obtained when we set $\omega=\Omega$  and $\delta=\pi$.
The dotted line represents the Lorentz factor in Eq.~(\ref{eq:integ1}), 
$[1-A^2  \sin^2(\Omega \, t)]^{-1/2}$, 
while the solid line represents the cosine term, $-\cos(2 \,\Omega \, t - \delta) =
\cos(2 \,\Omega \, t )$ both graphed as 
functions of time. Here we have chosen $A=0.2$ and $\Omega =2$. (The figures have been appropriately scaled to allow easier visualization.)}
\label{fig:perpfig1pw} 
\end{figure}

The integrated energy density along the accelerated observer's worldline is 
\begin{equation}
I = \int \langle T^{\mu \nu} u_{\mu} u_{\nu} \rangle d\tau \,,
\label{eq:I}
\end{equation}
where the integrand is
\begin{equation}
\langle T^{\mu \nu} u_{\mu} u_{\nu} \rangle  d\tau =
\gamma^2 \, \langle T^{tt} \rangle d\tau 
=  \langle T^{tt} \rangle \, \frac{dt}{\sqrt{1-A^2 {\rm sin}^2(\omega t)}} \,.
\label{eq:Tmunu}
\end{equation} 
Here we used the facts that $\langle T^{tx} \rangle =\langle T^{xx} \rangle=0$
and  $\gamma^2 d\tau = \gamma \, dt $. If we expand to first order in $r$, the result
is 
\begin{equation}
\langle T^{\mu \nu} u_{\mu} u_{\nu} \rangle  d\tau \approx 
 -\frac{r \, \Omega \, \cos(2 \,\Omega \, t - \delta) \,dt}
{V \sqrt {1-A^2 {\rm sin}^2(\omega \, t)} } \,.
\label{eq:integ1}
\end{equation}
The numerator of this expression describes the fact that, for small squeeze parameter,
the inertial frame stress tensor components are nearly sinusoidal. The denominator describes
the effect of going to the non-inertial frame. If we can arrange that the $\gamma$ factor  
has its maximum value when the numerator is negative, then accelerated observer will
see net negative energy.  This situation occurs when $\omega = \Omega$ and when $\delta = \pi$, 
 as illustrated in Fig.~\ref{fig:perpfig1pw}. We will make this choice 
 throughout the remainder of this subsection.

In this case, the integrated energy density becomes
\begin{equation}
I = \frac{r \, \Omega}{V} \int \frac{dt}{\sqrt{1-A^2 {\rm sin}^2(\Omega t)}}  \,
  {\rm cos}(2 \,\Omega\, t ) \,. 
 \label{eq:I2}
 \end{equation}
 If we perform the integration  on $t$ and multiply by the quantization 
 volume, we get
\begin{equation}
I \, V\approx \frac{r \, [2 \, E(\Omega \,t, A^2) + (A^2 - 2) \, F(\Omega \, t, A^2)}{A^2}\,,
\label{eq:int3}
\end{equation}
where  $F(\Omega \, t, A^2)$ and $E(\Omega \, t, A^2)$ are elliptic integrals of the first  and 
second kind, respectively.

\begin{figure} 
\begin{center} 
\includegraphics[width=5in]{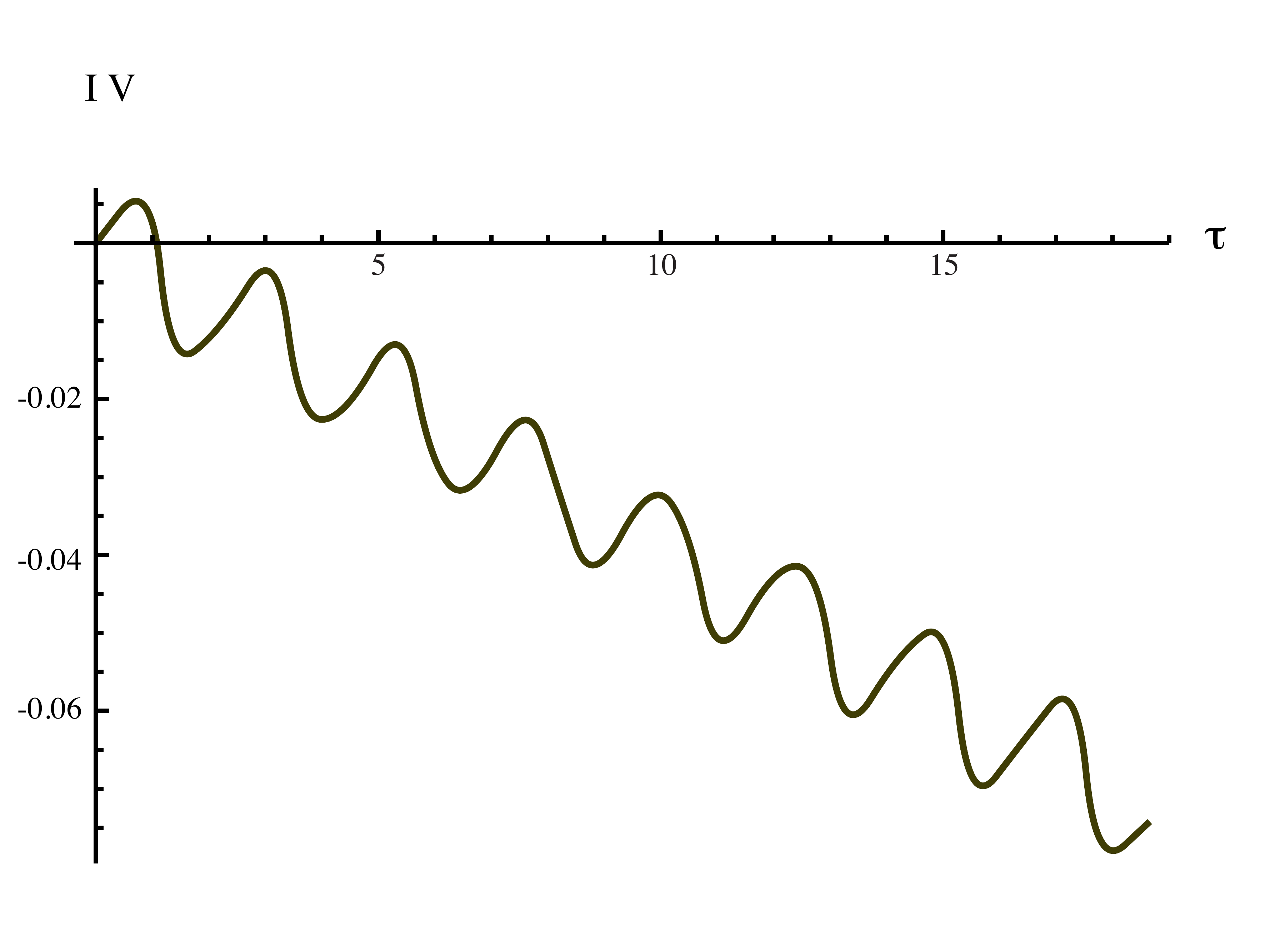}
\end{center} 
\vspace{-0.5in}
\caption{The integrated energy density multiplied by the quantization volume, 
$I \, V$,  seen by an accelerated observer who is moving 
perpendicularly to the direction of wave propagation, is plotted as a function 
of his proper time, $\tau$, for the parameters $ r=0.01$, $A = 0.9$, and in units 
where $\Omega = 1$.}  
\label{fig:oscpwFig2} 
\end{figure}

As a specific example, let us  plot $I \, V$ for 
$ r=0.01$, $A = 0.9$, and in units where $\Omega = 1$.  
Since, strictly speaking, the energy density is inversely proportional to $V$,  we want 
to make a graph of $I \,V$ as a function of $\tau$, i.e., a graph of the integrated energy density, 
multiplied by the quantization volume, seen by the accelerated observer as a function 
of his proper time. The relation between $\tau$ and $t$ is $\tau = \int dt/ \gamma$, which is 
\begin{equation}
\tau = \frac{E (\Omega \, t,A^2)}{\Omega} \,.
\label{eq:tau}
\end{equation}
If we plot Eq.~(\ref{eq:int3}) against Eq.~(\ref{eq:tau}) for our chosen parameters, we get 
Fig.~\ref{fig:oscpwFig2}.

Now let us examine our expression for $I$ in the $A\ll 1$ limit. If 
we expand the Lorentz factor to second-order in $A$, we obtain
\begin{equation}
\frac{1}{\sqrt{1-A^2 {\rm sin}^2(\Omega t)}} \approx  1+ \frac{1}{2} A^2  {\rm sin}^2(\Omega\,  t) +
 {\mathcal O}(A^4) \,.
 \label{eq:gampar}
\end{equation}
In this limit, the difference between $dt$ and $d \tau$ will be $ {\mathcal O}(A^2)$. 
If we use Eq.~(\ref{eq:gampar}) in Eq.~(\ref{eq:I2}) to calculate $I$, we find:
\begin{equation}
I \approx -\frac{r A^2 \, \Omega \, T}{8 V} 
+ \frac{r \, {\rm sin}(2 \, \Omega \, T)}{2 V} + 
 \frac{r \, A^2 {\rm sin}(2\, \Omega \, T)}{8 V} -\frac{r \, A^2 {\rm sin}(4 \, \Omega \, T)}{32 V} \,.
\end{equation}
The sinusoidal terms will eventually be dominated by the linear term, but this can take many 
cycles, so we keep the ${\mathcal O}(A^0)$ sinusoidal term, but drop the 
 ${\mathcal O}(A^2)$ sinusoidal terms. Therefore, our two leading order terms are 
 \begin{equation}
I \approx -\frac{r A^2 \, \Omega \, T}{8 V}  + \frac{r \, {\rm sin}(2 \, \Omega \, T)}{2 V}  \,.
\label{eq:smallA}
\end{equation}
Here $A^2 \ll 1$, so the oscillating term is larger in magnitude until $ T >  4 /  (A^2 \, \Omega)$. 
After this, the linear term dominates. However, we should recall that there is positive $r^2$ term
in Eq.~(\ref{eq:small_r}). This term will give a contribution to $I$ of $\Omega\,r^2\,T/V$, and is
negligible only if we require that
\begin{equation}
 8\, r \ll A^2\,. 
\label{eq:Alim}
\end{equation}
Nonetheless, accumulating negative energy density, can occur for 
arbitrarily small velocities. For any $A\not=0$, we can find a value of $r$ which satisfies
Eq.~(\ref{eq:Alim}). Then eventually the first term in  Eq.~(\ref{eq:smallA})
will dominate.

\subsection{Parallel Motion}
\label{sec:parmotion}

We now consider the case of the accelerating observer moving parallel to the direction 
of the propagation of the wave. In the lab frame, we have 
$\langle T^{tt} \rangle = \langle T^{zz} \rangle = \langle T^{tz} \rangle$. 
The accelerated observer's three-velocity and position, respectively, are
\begin{eqnarray}
v&=& v^z(t) = \frac{dz}{dt}  = A \,{\rm sin}(\omega\, t) \,, 
\label{eq:vpar} \\
z(t) &=& -\frac{A}{\omega}\,{\rm cos}(\omega \,t)
\label{eq:zpar}
\end{eqnarray}
and so
\begin{equation}
u^{\mu}=\gamma (1,0,0,v) \,,
\end{equation}
where $u^t = \gamma ={dt}/{d\tau}$. Therefore, we have that
\begin{equation}
\langle T^{t't'} \rangle = \gamma^2 
\,\Bigl(1-2v+v^2 \Bigr)\, \langle T^{tt} \rangle 
= \Biggl(\frac{1-v}{1+v} \Biggr)\, \langle T^{tt} \rangle \,
\end{equation}
where $(1-v)/(1+v)$ is a linear Doppler shift factor (as opposed to 
the transverse Doppler factor in the perpendicular case).
As a result,
\begin{equation}
I =  \int \langle T^{\mu \nu}\rangle u_{\mu} u_{\nu} d\tau 
= \int \Biggl(\frac{1-v}{1+v} \Biggr)\, \langle T^{tt} \rangle\, d\tau  
=\int \frac{\,\,(1-v)^{3/2}}{\sqrt{1+v}}\, \langle T^{tt} \rangle\, dt \,,
\label{eq:int4}
\end{equation}
since $d\tau = \gamma^{-1} dt = \sqrt {1- v^2} dt$.

\begin{figure} 
\begin{center} 
\includegraphics[width=5in]{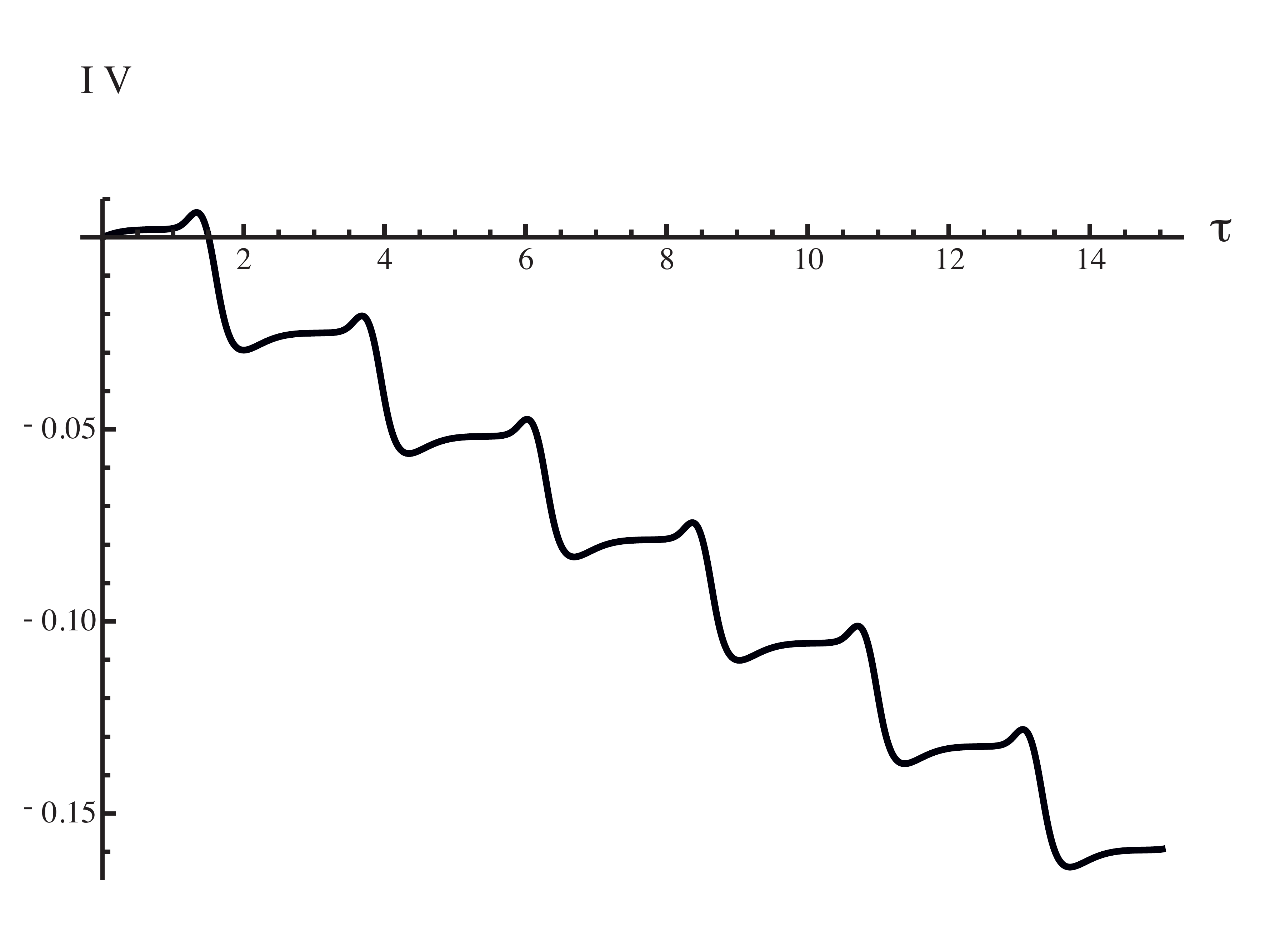}
\end{center} 
\vspace{-0.5in}
\caption{The integrated energy density multiplied by the quantization volume, 
$I \, V$,  seen by an accelerated observer who is moving 
parallel to the direction of wave propagation, is plotted as a function 
of his proper time, $\tau$, for the parameters $ r=0.01$, $A = 0.9$, and 
$\omega = 2$, in units 
where $\Omega = 1$. Note that the accumulated negative energy density grows 
much faster than in the case of the perpendicularly  moving observer, due 
to the linear Doppler shift term in the energy density. We also see extra structure 
in this curve as well, because the expression for the energy density is more 
complicated than in the perpendicular motion case.}  
\label{fig:oscpwFig3} 
\end{figure}

Here the observer is moving in the direction of wave propagation, so we can no longer set $z=0$.
Now the energy density in the inertial frame is given by Eq.~(\ref{eq:Exsq}),  with 
$z = -(A/ \omega)\, {\rm cos}(\omega \,t)$, so
\begin{equation}
\langle T^{tt} \rangle = \frac{\Omega}{V} \, \sinh r \, \left\{\sinh r -\cosh r \, 
\cos \left[\left(\frac{2 A \Omega}{\omega} \right) \cos(\omega t) + 2\Omega t -\delta \right] \right\} \,.
\label{eq:Tinertial}
\end{equation}
In this case, we find accumulating negative energy density for $\omega = 2 \, \Omega$, and 
$\delta = -\pi/2$.  The integral in Eq.~(\ref{eq:int4}) can be done analytically for small $A$, as will be discussed
 below, but for more general $A$, it can only be performed numerically.
As an example, let us choose the case where  $r = 0.01$,  $\delta =- \pi/2$, $A = 0.9$, and $\omega = 2$, 
in units where $\Omega=1$. In Fig.~\ref{fig:oscpwFig3}, we graph $ I V$ against the observer's proper time, which 
will again be given by Eq.~(\ref{eq:tau}).

Now we wish to consider the non-relativistic limit, and work to first order
in $v$ and hence in $A$. To this order, $d \tau \approx dt$, so 
\begin{equation}
I  \approx \int  \langle T^{t't'} \rangle \, dt\,,
\label{eq:intpart}
\end{equation}
where the energy density in the accelerating frame is
\begin{equation}
\langle T^{t't'} \rangle \approx (1-2v)\, \langle T^{tt} \rangle \,.
\end{equation}
If we expand Eq.~(\ref{eq:Tinertial}) to first order in $A$, the result is
\begin{equation}
\langle T^{tt} \rangle \approx \frac{\Omega}{V} \, \sinh r \, \left\{\sinh r -\cosh r \, 
\left[\cos(2\Omega t -\delta) - 
\frac{2 \Omega}{\omega} \, A \, \cos(\omega t) \,
\sin(2\Omega t -\delta)  \right] \right\} \,,
\label{eq:Tinertial2}
\end{equation}
where we have used the fact that 
\begin{eqnarray}
\cos\Biggl[\Biggl(\frac{2 \Omega \,A}{\omega} \Biggr) \, \cos(\omega t) \Biggr]
&\approx& 1 + O(A^2) \,,
\nonumber \\
\sin\Biggl[\Biggl(\frac{2 \Omega \,A}{\omega} \Biggr) \,\cos(\omega t)\Biggr]
&\approx& \Biggl(\frac{2 \Omega \,A}{\omega} \Biggr) \,\cos(\omega t) + O(A^3) \,.
\label{eq:cossinapprox}
\end{eqnarray}
Next we evaluate the energy density in the accelerating frame to first order in
$A$ and set $\omega = 2\Omega$ to find
\begin{eqnarray}
\langle T^{t't'} \rangle &\approx&  \frac{\Omega}{V} \, \sinh r \,
 \left\{ \sinh r \,\left[1 -2A \sin(2\Omega t) \right]  \right. \nonumber \\
 &-&\left. \cosh r \, \left[\cos(2\Omega t -\delta) -\frac{3}{2} A \sin(4\Omega t -\delta) 
- \frac{1}{2} \, A\, \sin \delta \right] \right\}\,.
\end{eqnarray}
This expression reveals that we can have growing negative energy density if $\delta = -\pi/2$
and $r \ll 1$. In this case, we may write
\begin{equation}
\langle T^{t't'} \rangle \approx  \frac{\Omega}{V} \,\left[ r^2 
-r \left(\frac{1}{2}\, A -\sin(2\Omega t) \right) \right] \,,
\label{eq:rho_acc}
\end{equation}
where order $A$ oscillatory terms have been dropped. If  
\begin{equation}
r \ll \frac{1}{2}\, A\,,
\label{rlimpar}
\end{equation}
which is the analog of Eq.~(\ref{eq:Alim}),  
the integrated energy density grows negatively as 
\begin{equation}
I \approx -\frac{r \,A \, \Omega \, T}{2 V}  + \frac{r}{2 V} [1 - \cos(2 \Omega T)]
\sim -\frac{r \,A \, \Omega \, T}{2 V} \,.
\label{eq:rate}
\end{equation}
The latter asymptotic form holds for 
\begin{equation}
T \gg \frac{1}{\Omega\, A}\,.
\end{equation}
In the parallel motion case, the rate of growth of the negative integrated energy density
is first order in $A$, as compared to second order in the perpendicular motion case
treated in the previous subsection. 
This is due to the fact that in the parallel case, there is a linear 
Doppler shift, whereas in the perpendicular case the Doppler shift is transverse.

\section{Oscillations in a Cavity}
\label{sec:cavity}
\subsection{The Perpendicular Case}
\label{sec:perpcav}
We now consider the case of a particle oscillating in a closed cavity 
with dimensions $a$, $b$, and $d$ aligned along the 
$x$, $y$, $z$ axes respectively, where $b < a < d$. 
The modes in this cavity were discussed in 
Ref.~\cite{FR09}. With the condition that $b<a<d$, the 
lowest frequency mode is the TE mode with $p=l=1,m=0$, where 
the frequency of the mode is given by
\begin{equation}
\Omega = \pi \sqrt{\frac{1}{a^2}+\frac{1}{d^2}} \,,
\label{eq:lowestmode}
\end{equation}
and the non-zero components of the electric and magnetic fields are
\begin{eqnarray}
{\cal E}_x &=&  {\cal E}_z=0  \,,\nonumber\\
{\cal E}_y&=& \frac{ \Omega a}{\pi} \, C \, \sin \Big(\frac{\pi}{a} x \Big)\, 
\sin  \Big(\frac{\pi}{d} z \Big) \, e^{-i \, \Omega\, t} \,, \nonumber \\
{\cal B}_x &=& i \,\,\frac{a}{d} \,\, C\, \sin \Big(\frac{\pi}{a} x \Big)\, 
 \cos \Big(\frac{\pi}{d} z \Big) \, e^{-i \, \Omega\, t} \,, \nonumber \\
{\cal B}_y &=& 0 \,, \nonumber \\
{\cal B}_z &=& -i \, C \,\cos \Big(\frac{\pi}{a} x \Big) \, \sin \Big(\frac{\pi}{d} z \Big)
\, e^{-i \, \Omega\, t} \,, 
\label{eq:EBs}
\end{eqnarray}
where the electric field is taken to be polarized in the
$y$-direction. This mode is independent of $y$. 
Here $C$ is a real normalization constant, given by
\begin{equation}
C^2 = \frac{2 \, \Omega}{a\,b\,d\, (1+a^2/d^2) } \,.
\label{eq:C}
\end{equation}

For the case where only a single mode $j$ is excited,  the normal ordered 
expectation values of the squared fields are
\begin{equation}
\langle E^2 \rangle = 2 \langle a^\dagger \, a\rangle  \, |{ \cal E}_j|^2+ 
2Re \Big( \langle  a^2 \rangle \, { \cal E}_j^2 \Big) 
\end{equation}
and 
\begin{equation}
\langle B^2 \rangle = 2 \langle a^\dagger \, a\rangle  \, |{ \cal
  B}_j|^2 + 
2Re \Big( \langle  a^2 \rangle \, {\cal B}_j^2 \Big)\,.
\end{equation}
where
\begin{eqnarray}
 {\cal E}_j^2& =& {{\cal E}_y}^2 =  \frac{\Omega^2 a^2}{\pi^2} \, C^2 \, 
 {\rm sin}^2  \Big(\frac{\pi}{a}x \Big)\, {\rm sin}^2
 \Big(\frac{\pi}{d}z \Big)\, e^{-2i\Omega t}
\nonumber \\
 {\cal B}_j^2 &=& {\cal B}_x^2 + {\cal B}_z^2 
= - C^2 \Bigg[{\rm cos}^2 \Big(\frac{\pi}{a}x \Big) \, {\rm sin}^2 \Big(\frac{\pi}{d}z \Big)
+ \frac{a^2}{d^2} \, {\rm sin}^2 \Big(\frac{\pi}{a}x \Big)\,
{\rm cos}^2 \Big(\frac{\pi}{d}z \Big)   \Bigg]\, e^{-2i\Omega t} \,.
\end{eqnarray}

In this case, we can have the particle moving in the $y$-direction, and located 
in the center of the cavity in the other directions, so that
\begin{equation}
x=\frac{a}{2},\, z=\frac{d}{2} \,.
\end{equation} 
This considerably simplifies the mode functions, leading to
\begin{equation}
{\cal B}_x = {\cal B}_z =0\,, \qquad {\cal E}_y = \frac{\Omega \, a}{\pi} \,C\, {\rm e}^{-i \Omega t}\,.
\end{equation}
The only non-zero components of the stress tensor which we will need are 
$\langle T^{tt} \rangle$
and $\langle T^{yy} \rangle$, which become
\begin{eqnarray}
\langle T^{tt} \rangle &=& -\langle T^{yy} \rangle = 
\sinh r \left[\sinh r \, |{\cal E}_y|^2 -\cosh r \,{\rm Re}\left({\rm e}^{i\delta} \,  {\cal E}_y^2 \right) \right]
\nonumber \\
&=& N \, \sinh r \left[\sinh r - \cosh r \, \cos(2 \Omega t -\delta) \right]\,,
\label{eq:ttyy}
\end{eqnarray}
where
\begin{equation}
N=\left(\frac{\Omega \, a}{\pi} \, C\right)^2 \,.
\label{eq:N}
\end{equation}

Because the direction of oscillation of the particle is in the $y$-direction, 
\begin{equation}
u^{\mu}=\gamma (1, 0, v^y, 0) \,,
\end{equation}
where $v^y=A\, \sin \omega t$. The integrand of $I$ is 
\begin{equation}
\langle T_{\mu \nu}  u^{\mu} u^{\nu}\rangle d\tau 
=  \frac{(\langle T_{tt} \rangle +({v^y})^2 \langle T_{yy} \rangle) }
{\sqrt{1-A^2 {\rm sin}^2(\omega t)}} \, dt 
= \sqrt{1-A^2 {\rm sin}^2(\omega t)} \, \langle T_{tt} \rangle \, dt \,. 
\label{eq:integ5}
\end{equation}

Let
\begin{equation}
U = \sqrt{1-A^2 {\rm sin}^2(\omega t)} \,\langle T_{tt} \rangle \,. 
\end{equation}
We will assume that $A \ll 1$, and expand to second order in $A$. 
Therefore, if we use Eq.~(\ref{eq:ttyy}) in Eq.~(\ref{eq:integ5}), 
and set $\omega=\Omega$, we have that 
\begin{equation}
 U \approx   N \,\Bigl(1- \frac{1}{2} \, A^2 \sin^2(\Omega t) \Bigr) \,
[\sinh^2 r - \sinh r\,\cosh r \cos(2 \Omega t - \delta) ]  \,.
\label{eq:URHS}
\end{equation}
 If we now expand the right-hand side of Eq.~(\ref{eq:URHS}) to second order in $r$, 
set $\delta = 0$, integrate from 
$0$ to $T$, and drop oscillatory terms in $A^2$, we obtain
\begin{equation}
I \approx N \, \Biggl( r^2 T - \frac{1}{8} \, A^2 r T - r \, \frac{\sin(2 \Omega T)}{2 \, \Omega} \Biggr) \,,
\end{equation}
where we have also dropped a higher order $A^2 \, r^2$ term.

The positive first term is negligible compared to the second when 
\begin{equation}
r \ll  \frac{1}{8} \, A^2 \,.
\end{equation}
The middle negative linear growing term will dominate the sinusoidal term 
when 
\begin{equation}
T > \frac{4}{A^2 \, \Omega}
\end{equation}
In this case, the restrictions on $r$ and $T$ are the same as 
those for perpendicular motion in the plane wave case. 
In these limits, we therefore have negative energy density 
which grows linearly as
\begin{equation}
I \approx   - \frac{1}{8} \, N\, A^2 r T  \,.
\end{equation}
If we use Eqs.~(\ref{eq:C}) and (\ref{eq:N}), we can write the previous equation as 
\begin{equation}
I \approx   - \frac{r \, A^2 \, \Omega \, T }{4 V}  \,,
\end{equation}
where $V=abd$ is the volume of the cavity. Compare this result with 
the first term in Eq.~(\ref{eq:smallA}), the
corresponding rate for perpendicular motion in a plane wave mode. If we identify the cavity volume
in the former with the quantization volume in the latter, then they differ only by a factor
of two.

\subsection{The Parallel Case}
\label{sec:parcav}
In this subsection, we will consider the case of a particle oscillating in a cavity along 
 the $z$-axis, in the limit where $A \ll 1$, and work to first order in $A$. We take 
 \begin{equation}
v_z=A\, {\sin\,}\omega t \,,
\end{equation}
and 
 \begin{equation}
z=z_0 - \frac{A\, {\cos} \, \omega t }{\omega} + O(A^2) \,,
\label{eq:z-dep}
\end{equation}
where $z=z_0$ corresponds to the equilibrium position of the particle, and 
the last term corresponds to relativistic corrections.
We will choose the $x$-position of the particle to be
\begin{equation}
x=\frac{a}{2} \,.
\end{equation}
 The energy density in the particle's frame is
\begin{equation}
\langle T^{t' t'} \rangle =\gamma^2 \,
\langle (T^{tt}-2 v_z \, T^{tz} + {v_z}^2 \, T^{zz}) \rangle \approx 
\langle T^{tt}\rangle  -2 v_z \,\langle  T^{tz} \rangle   \,.
\label{eq:Ttt'}
\end{equation}
Here 
\begin{equation}
\langle T^{tz} \rangle = -\langle E_y \, B_x \rangle \,.
\end{equation}
We need to calculate $\langle T^{tt}\rangle$ and $\langle T^{tz}\rangle$
using the mode functions in Eq.~(\ref{eq:EBs}), and then expand the result to
second order in $r$. The result for $\langle T^{t' t'} \rangle$
may be written as
\begin{equation}
\langle T^{t' t'} \rangle \approx C^2\, \{ (F_1 +2 v_z\, F_3) r^2 - 
r\,[F_2 \, \cos(2 \Omega t -\delta) +2 v_z\, F_3\, \sin(2 \Omega t -\delta)]\}\,,
\label{eq:Tparcav}
\end{equation}
where
\begin{equation}
F_1= F_1(z) =\frac{\Omega^2 \,a^2}{\pi^2} \,\Biggl[{\rm sin}^2 \Biggl(\frac{\pi z}{d} \Biggr)
+\frac{a^2}{d^2} \,{\rm cos}^2 \Biggl(\frac{\pi z}{d} \Biggr)\Biggr] \,,
\label{eq:F1}
\end{equation}
\begin{equation}
F_2= F_2(z) =\frac{\Omega^2 \,a^2}{\pi^2} \,\Biggl[{\rm sin}^2 \Biggl(\frac{\pi z}{d} \Biggr)
-\frac{a^2}{d^2} \,{\rm cos}^2 \Biggl(\frac{\pi z}{d} \Biggr)\Biggr] \,,
\label{eq:F2}
\end{equation}
and
\begin{equation}
F_3 = F_3(z) = \frac{\Omega\,a^2}{\pi d} \,
{\rm sin}\Biggl(\frac{2\,\pi z}{d}\Biggr) \,,
\label{eq:F3}
\end{equation}
and where $C^2$ is once again given by Eq.~(\ref{eq:C}).

The integrated energy density may be written as
\begin{equation}
I \approx \int dt\, \langle T^{t' t'} \rangle \approx C^2\, ( r^2\, I_1 -r \, I_2)
\end{equation}
where
\begin{equation}
I_1 = \int_0^T \, dt \, (F_1 +2 v_z\, F_3)\,,
\end{equation}
and   
\begin{equation}
I_2 = \int_0^T \, dt \, [F_2 \, \cos(2 \Omega t -\delta) +2 v_z\, F_3\, \sin(2 \Omega t -\delta)]\,.
\end{equation}
As in the case of parallel motion in the plane wave case, 
with the appropriate choices for $\omega, \, \Omega,\, r$ and $\delta$, 
we expect to get a linearly growing negative term, a term which is first order in $r$ 
and sinusoidal in time, and a positive $r^2$ term. The first 
and second of these terms will arise from $F_2$ and $F_3$, 
while the third term will arise from $F_1$.   We also expect that
we will find a non-trivial effect in first order in $A$.

Let us first examine the terms involving $F_3$ in Eq.~(\ref{eq:Tparcav}). 
These terms both involve the product $v_z\,F_3$, and are hence already
of order $A$. Thus we may use Eq.~(\ref{eq:z-dep}) to write
\begin{equation}
F_3(z) \approx F_3(z_0) = \frac{\Omega \,a^2}{\pi \,d} \,,
\end{equation}
where we have set $z_0=d/4$. (As it turns out, the linearly growing term we want will 
come from the $F_3$ term in $I_2$, so we cannot choose $z_0=d/2$.)

A similar situation applies to $F_1$, which contributes only to an order
$r^2$ term. This is a positive, growing term which we need only to
zeroth order in $A$. For this purpose, we may evaluate $F_1$
at $z=z_0$:
\begin{equation}
F_1(z) \approx F_1(z_0) = 
  \frac{\Omega^2 \,a^2}{2 \pi^2} \, \left(1 +\frac{a^2}{d^2} \right) \,.
\label{eq:F1approx}
\end{equation} 
Thus, for estimating the order $r^2$ term, we may use
\begin{equation}
I_1 \sim F_1\, T\,,
\end{equation}
where $F_1$ has the value in Eq.~(\ref{eq:F1approx}). 

The negatively growing term comes from
$I_2$, which involves $F_2$, so we need to expand the latter to first order in 
$A$, using Eq.~(\ref{eq:z-dep}), as
\begin{eqnarray}
F_2(z) &=& \frac{\Omega^2 \,a^2}{2 \pi^2} \,\left[ \left(1 - \frac{a^2}{d^2} \right) -
\left(1 + \frac{a^2}{d^2} \right)\, \cos\left(\frac{2 \pi z}{d}\right) \right] \nonumber \\
&\approx&  \frac{\Omega^2 \,a^2}{2 \pi^2} \,\left[ \left(1 - \frac{a^2}{d^2} \right) -
\left(1 + \frac{a^2}{d^2} \right)\, \left(\frac{2 \pi A}{d\, \omega}\right)\, \cos(\omega t) \right] \,.
\end{eqnarray}
The $I_2$ term will be maximally negative when $\omega = 2\, \Omega$ and $\delta =0$.
In this case, a short calculation yields
\begin{equation}
I_2 \approx A\,  \frac{\Omega \,a^2}{4 \pi d} \, \left(3 - \frac{a^2}{d^2} \right) \, T
+ \frac{\Omega \,a^2}{4 \pi } \, \left(1 - \frac{a^2}{d^2} \right) \, \sin(2 \Omega T) \,,
\end{equation}
where oscillatory, order $A$ terms have been dropped. Note that $3 - {a^2}/{d^2} > 0$ because
$a < d$. 

Therefore, the integrated energy density becomes,
\begin{equation}
I \approx C^2 \, \left[ 
- r \, A\,T\,  \frac{\Omega \,a^2}{4 \pi d} \, \left(3 - \frac{a^2}{d^2} \right)  
-  \,r\, \frac{\Omega \,a^2}{4 \pi^2 } \, \left(1 - \frac{a^2}{d^2} \right) \, \sin(2 \Omega T) 
+r^2 \, T\, \frac{\Omega^2 \,a^2}{2 \pi^2} \, \left(1 + \frac{a^2}{d^2} \right) \right]\,.
\end{equation}
We see that the negative linearly growing term will dominate 
the sinusoidal term when 
\begin{equation}
T > \frac{d}{\pi A}\, \left(\frac{d^2-a^2}{3d^2-a^2} \right) \,.
\end{equation}
and the positive, order $r^2$ term, when
\begin{equation}
 \frac{2 \Omega d}{\pi}\, \left(\frac{d^2+a^2}{3d^2-a^2} \right) \, r < A \,.
\label{rlimparcav}
\end{equation}
In this case, we find that the integrated energy density in the particle's frame grows negatively as
\begin{equation}
I \sim - \frac{r A \Omega}{2 V} \, T \;
 \left[\frac{\Omega \,a^2}{\pi d } \, \Biggl(\frac{3d^2-a^2}{d^2+a^2} \Biggr) \right]\,,
\end{equation}
where we have used the definition of $C^2$ and the fact that $V= a b d$ is the volume of the cavity. Compare this result with Eq.~(\ref{eq:rate}), the
corresponding rate for parallel motion in a plane wave mode. If we identify the cavity volume in the former with the quantization volume in the latter, then they differ only by the factor in the square brackets. If $a$ and $d$ are of the same order of magnitude, then Eq.~(\ref{eq:lowestmode}) tells us that $\Omega \sim O(1/a) \sim O(1/d)$, and this factor is of order unity.

\section{Effects of the Negative Energy on Focussing}
\label{sec:focus}

In this section, we will treat one possible effect of the accumulating negative energy along a particle's
worldline. It is well-known that the attractive character of gravity, with ordinary matter as a source,
leads to focussing of null and timelike geodesics. One expects that negative energy densities might
have the opposite effect, and produce defocussing through repulsive gravitational effects. 

\subsection{Raychaudhuri Equation}

The effect of
gravity on a congruence of timelike worldlines is described by the Raychaudhuri equation. In our case,
we allow the worldlines to be non-geodesics, so the equation takes the form~\cite{HE}
\begin{equation}
{\dot \theta} = \frac{d\theta}{d\tau} = 
-R_{\alpha \beta} u^{\alpha} u^{\beta} + 2\omega_{\alpha \beta }\omega^{\alpha \beta }  
- 2\sigma_{\alpha \beta }\sigma^{\alpha \beta }  -\frac{1}{3}\theta^2+ \nabla_{\beta} a^{\beta}\,.
\label{eq:ray}
\end{equation} 
Here  $u^\alpha$ and $a^\beta = u^\alpha \nabla_\alpha u^\beta$ are the 4-velocity and 4-acceleration
of the congruence,  and $\sigma_{\alpha \beta }$ and $\omega_{\alpha \beta }$ are the shear and vorticity
tensors. Also, $\theta = \nabla_\alpha u^\alpha$ is the expansion, and  $R_{\alpha \beta }$ is the Ricci
tensor. The last term in Eq.~(\ref{eq:ray}) is the acceleration term, which vanishes for geodesics.
We will assume a hypersurface orthogonal congruence, in which case the vorticity tensor vanishes,
$\omega_{\alpha \beta } =0$. In addition, we assume that the shear and expansion are sufficiently small,
that the terms quadratic in those quantities may be neglected. In this case, the Raychaudhuri equation
becomes
\begin{equation}
{\dot \theta} \approx  -R_{\alpha \beta} u^{\alpha} u^{\beta} + {\dot \theta}_{\rm ac} \,,
\label{eq:ray2}
\end{equation} 
where $ {\dot \theta}_{\rm ac} = \nabla_{\beta} a^{\beta}$ is the acceleration term, and the Ricci tensor term
describes the effects of gravity.

Next we assume that an electromagnetic field is both the cause of the acceleration and the sole source
of the gravitational field. Particles with rest mass $m$ and electric charge $q$ obey the equation of motion
\begin{equation}
a_{\beta}=\frac{q}{m} \,F_{\beta \rho} \, u^{\rho} \,,
\label{eq:acbeta}
\end{equation}
where the field strength tensor, $F_{\beta \rho}$, is assumed to obey the source free equation
\begin{equation}
\nabla_{\alpha}\,F^{\alpha \beta} =0 \,.
\end{equation}
We can now write the acceleration term as
\begin{equation}
{\dot \theta}_{\rm ac} = \frac{q}{m}F^{\alpha \beta}(\nabla_{\alpha} u_{\beta}) \,.
\label{eq:dottheta}
\end{equation}
The covariant derivative of the 4-velocity may be expressed as~\cite{MTW} 
\begin{equation}
\nabla_{\alpha} u_{\beta}= \sigma_{\beta \alpha}
+\frac{1}{2} \theta (g_{\alpha \beta}+u_{\alpha} u_{\beta})- a_{\beta} u_{\alpha} \,,
\label{eq:delalphaubeta}
\end{equation}
when $\omega_{\beta \alpha} =0$. However, all terms on right hand side of this expression, except for
the last, are symmetric tensors which vanish when contracted into the antisymmetric field strength tensor.
Thus we obtain
\begin{equation}
{\dot \theta}_{\rm ac} = -a_{\beta} a^{\beta}\,.
\label{eq:thetaac}
\end{equation}

The electromagnetic stress tensor, given in Eq.~(\ref{eq:T}) is tracefree, so the Einstein equations become
\begin{equation}
R_{\alpha \beta}
=8 \pi {\ell_p}^2 \, T_{\alpha \beta} \,,
\label{eq:Rab}
\end{equation}
where $\ell_p$ is the Planck length, and Newton's constant is $G= {\ell_p}^2$, in units where $\hbar = c =1$.
We may write
\begin{equation}
T_{\alpha \beta} u^{\alpha} u^{\beta} = 
(u^{\alpha}F_{\alpha \rho}) u^{\beta} {F_{\beta}}^ {\rho} 
+\frac{1}{4}F_{\mu \nu}F^{\mu \nu} 
=\frac{m^2}{q^2} \, a_{\rho} a^{\rho} + \frac{1}{4}F_{\mu \nu}F^{\mu \nu} \,,
\label{eq:Tab}
\end{equation}
where we have used Eq.~(\ref{eq:acbeta}). We may use this expression to evaluate the Ricci tensor term
in the Raychaudhuri equation and write
\begin{equation}
{\dot \theta} \approx -\Biggl( 1+ 8 \pi \frac{{\ell_p}^2 \, m^2}{q^2} \Biggr) \,
 a_{\rho} a^{\rho} - 2  \pi {\ell_p}^2 \, F_{\mu \nu}F^{\mu \nu} \,.
 \label{eq:dottheta2}
 \end{equation}

 \subsection{Fields Producing Acceleration}
 
 In previous sections, we assumed a prescribed sinusoidal motion, but did not explicitly give the electromagnetic
 fields which would produce this motion. Here we will concentrate on the case of motion parallel to a plane 
 wave mode, which was treated in Sec.~\ref{sec:parmotion}.
 In particular, we consider the case of non-relativistic motion along the
 $z$-direction, as described by Eq.~(\ref{eq:vpar}), with $A \ll 1$.  This motion can approximately be produced
 by a plane wave with polarization in the $z$-direction. Here we will consider a classical electromagnetic wave
 propagating in the $y$-direction, with electric field ${\bf E}= E_c\, {\bf \hat{z}}$, and magnetic field
${\bf B}= E_c\, {\bf \hat{x}}$, where
\begin{equation}
E_c = \frac{\omega\, A \, m} {q}\; \cos\omega (t-y) \,. 
\label{eq:Ec}
\end{equation}
To order $A$, only the electric field determines the motion of the particle, with the magnetic force contributing
 in order $A^2$. Because the motion of the particle 
 is only in the   $z$-direction, we may set $y=0$. In this case, the 
energy density of the classical wave, in the laboratory frame, is
\begin{equation}
T_c^{tt} = E_c^2 = \left(\frac{\omega\, A \, m}{q}\right)^2 \; \cos^2(\omega t) \,. 
\label{eq:Tclass}
\end{equation}

In addition to this classical field, the particle is also subjected to the quantum fields associated with the
squeezed vacuum state mode. These fields potentially produce a fluctuating force on the particle, which
we wish to include. Let $\mathbf{E}_q$ and $\mathbf{B}_q$ be the terms in Eqs.~(\ref{eq:E-op}) and
(\ref{eq:B-op}), respectively, which refer to the mode in a squeezed vacuum state. That is,
\begin{equation}
\mathbf{E}_q = \hat{\bf x} \, ( { \mathcal E} \, a + { \mathcal E}^\ast \, a^\dagger)\,,
\label{eq:Eq}
\end{equation}
 and
 \begin{equation}
\mathbf{B}_q = \hat{\bf y} \, ( { \mathcal E} \, a + { \mathcal E}^\ast \, a^\dagger)\,,
\end{equation}
where $ { \mathcal E}$ is defined in Eq.~(\ref{eq:mode}). We will treat the velocity of the particle due
to the quantum electric field as an operator in the photon state space, $\mathbf{v}_q = v_q\,  \hat{\bf x}$,
where $v_q$ will be evaluated explicitly below. 

There is a third effect which we will not include explicitly. This the effect of the emitted radiation by
the particle. There will be an average radiation reaction force which will slightly change the
trajectory of the particle for a given classical field. However, this is normally very small and will
be neglected. There will also be a shot noise effect, an uncertainty in the particle's momentum
due to the statistical uncertainty in the number of photons emitted. This effect depends primarily
on the classical field driving the average motion and not upon the quantum electric field. Hence it, and the radiation reaction force, would cancel in any experiment 
which compares particle motion with and without the quantum 
electric field. In addition, this momentum uncertainty grows as the square root of the mean 
number of photons radiated, and hence as the square root of time. Here we are interested
in effects which grow linearly in time. 

We will now compute the components of the acceleration four-vector in the lab rest frame,
taking account of both the classical and quantum parts of the electromagnetic field and of the
particle's four-velocity. The  acceleration four-vector satisfies
\begin{equation}
a^\rho = \frac{q}{m}\, F^{\rho \alpha} \, u_\alpha \,.
\end{equation}
 In the non-relativisitic limit, the four-velocity is
 \begin{equation}
u^\alpha = (1, v_q,0,v_c)\,,
\end{equation}
where $v_c = A\, \sin (\omega t)$\,.
The non-zero components of the field strength tensor are 
\begin{equation}
F^{tz} = F^{yz} = E_c \qquad F^{tx} = F^{zx} = E_q\,,
\end{equation}
and those obtained by antisymmetry of $F^{\rho \alpha}$.
The components of $a^\rho$ become 
\begin{eqnarray}
a^t &=& \frac{q}{m}\, (E_q\, v_q + E_c \, v_c)   \nonumber \\
a^x &=& \frac{q}{m}\, E_q\, (1 - v_c)    \nonumber  \\
a^y &=& \frac{q}{m}\,  E_c \, v_c   \nonumber \\
a^z &=& \frac{q}{m}\, (E_q\, v_q + E_c ) \, .
 \label{eq:ax}
\end{eqnarray}
 
 We can now form the scalar $a^\rho a_\rho$, expand it to first order in the velocities, dropping $v_c^2$,
 $v_q^2$ and $v_c v_q$ terms, and take its expectation value in the squeezed vacuum state. The result
 is
 \begin{equation}
\langle a^\rho a_\rho \rangle \approx \frac{q^2}{m^2} \, \left[ \langle E_q^2 
\rangle (1- 2 v_c) + E_c^2
+2 E_c \langle E_q v_q \rangle \right] \,.
\label{eq:asq}
\end{equation}
Let us examine each term on the right-hand side of this expression. The classical energy density,
which is the same to first order in velocity  in the lab frame and in the particle rest frame, is just $ E_c^2$. 
Because the  classical wave is propagating in the $y$-direction, and 
all the motion is in the $x$ and $z$ directions, this is the perpendicular motion 
case, with respect to the classical wave. 
Thus, to order $v$, $T_c^{t't'} \approx T_c^{tt}$, 
since $T_c^{t't'} \approx T_c^{tt} + O(v^2)$.
The expectation value of the quantum energy density in the lab frame is $ \langle E_q^2 \rangle$, and 
is given explicitly by Eq.~(\ref{eq:Exsq}). This quantity in the particle rest frame is 
$ \langle E_q^2 \rangle (1- 2 v_c)$. The final term is the contribution of the velocity fluctuations to the
acceleration. 

For both the classical and quantum electromagnetic fields, we have assumed plane waves, for which
$E^2 = B^2$, and hence $ F_{\mu \nu}F^{\mu \nu} = 0$. Thus we may drop the last term in 
Eq.~(\ref{eq:dottheta2}), and write mean rate of change of the expansion as
\begin{equation}
\langle{\dot \theta}\rangle \approx -\Biggl( 1+ 8 \pi \frac{{\ell_p}^2 \, m^2}{q^2} 
\Biggr) \,\langle a_{\rho} a^{\rho} \rangle \,.
 \label{eq:dottheta3}
 \end{equation}

\subsection{Velocity Fluctuations and Defocussing}

The fluctuating part of the velocity, $v_q$, is determined by Eq.~(\ref{eq:ax}):
\begin{equation}
\frac{dv_q}{dt} = a^x =  \frac{q}{m}\, E_q\, (1 - v_c)\,,
\label{eq:ax2}
\end{equation}
where the term proportional to $v_c$ on the right hand side is due to the magnetic force produced
by $B_y$. Note that time derivative here is a total derivative, and we need to account for both the
explicit time dependence and the implicit dependence through $z(t)$:
\begin{equation}
\frac{dv_q}{dt} = \frac{\partial v_q}{\partial t} + \frac{\partial v_q}{\partial z} \, v_c \,,
\end{equation}
recalling that $v_c = dz/dt$. The solution to Eq.~(\ref{eq:ax2}) becomes
\begin{equation}
{v}_q = \frac{i q}{m \Omega}\, ( { \mathcal E} \, a - { \mathcal E}^\ast \, a^\dagger) \,,
\end{equation}
where we have used Eqs.~(\ref{eq:mode}) and (\ref{eq:Eq}). Note that the effects of the magnetic
force and of the implicit time dependence cancel one another.

We may compute $\langle E_q v_q \rangle = \langle :E_q v_q: \rangle$ in a squeezed vacuum state
to find
\begin{equation}
\langle E_q v_q \rangle = \frac{q}{m V}\, \sinh r \, \cosh r \, \sin[2\Omega (z-t) + \delta]\,.
\end{equation}
Here we used 
\begin{equation}
\langle a^2 \rangle = \langle (a^\dagger)^2 \rangle^\ast = -{\rm e}^{i\delta}\, \sinh r \, \cosh r  
\end{equation}
in the squeezed vacuum state. We may use Eq.~(\ref{eq:Ec}) with $y=0$, and set $\delta = -\pi/2$ to write
\begin{equation}
2 E_c\, \langle E_q v_q \rangle = -\frac{2 \omega\,A}{V} \, \sinh r \, \cosh r \, \cos(\omega t)\, \cos[2\Omega (z-t)] \,.
\end{equation}
We will work only to first order in $A$, which means that we can ignore the $z$-dependence (see Eqs.~(\ref{eq:zpar}) and ~(\ref{eq:cossinapprox})) in the above
expression, which will contribute in order $A^2$. When we set $\omega = 2 \Omega$, and drop oscillatory
terms, then we have
 \begin{equation}
2 E_c\, \langle E_q v_q \rangle \approx -\frac{2 \Omega\,A}{V} \, \sinh r \, \cosh r  \,.
\end{equation}
In the small $r$ limit, this becomes
\begin{equation}
2 E_c\, \langle E_q v_q \rangle \approx -\frac{2\Omega\,A}{V} \, r \,,
\label{eq:vfluct}
\end{equation}
which is to be compared with the same limit for the squeezed state energy density in the accelerated
frame,
\begin{equation}
\langle E_q^2 \rangle (1- 2 v_c) \approx  -\frac{ \Omega\,A}{2V} \, r \,.
\label{eq:negenergy}
\end{equation}
The latter quantity is just the order $r$, non-oscillatory term in Eq.~(\ref{eq:rho_acc}).
We see that both terms have the same form and same sign, and both contribute to defocussing,
although the effect of the quantum velocity fluctuations is four times that of the negative energy density
in this limit. 

If we combine these terms, as well as the time average of the classical energy density, 
Eq.~(\ref{eq:Tclass}), evaluated at $\omega = 2\Omega$, then Eq.~(\ref{eq:asq}) for the mean
squared acceleration becomes
 \begin{equation}
\langle a^\rho a_\rho \rangle \approx 2\, \Omega^2 \, A^2 - \frac{5 q^2\,\Omega\, A\, r}{4 V m^2} \,  \,.
\label{eq:asq2}
\end{equation}
The positive term is the focussing effect of the classical energy density, and the negative term
is the combined defocussing effect of the negative energy density and the velocity fluctuations.
These two terms depend upon different combinations of parameters, and it seems possible to
arrange for the defocussing effect to dominate.
Note that the gravitational effect, from the Ricci tensor, is $\propto {\ell_p}^2$ in 
Eq.~(\ref{eq:dottheta3}).  The part without $ {\ell_p}^2$ is a pure acceleration effect from the 
 acceleration term. However, both effects have the same functional form here.

 \section{Summary and Discussion}
 \label{sec:final}
 
 The key result of this paper is that an accelerated observer undergoing sinusoidal motion
 in space can observe an average constant negative  energy density, so the integrated energy density
 grows negatively in time in this observer's frame. This is contrast to an inertial observer, in whose frame
the energy density is more constrained by quantum inequalities. We considered a squeezed vacuum state
for both a plane wave and a standing wave in a cavity. The case in which 
growing integrated negative
energy is possible is when the squeeze parameter is small, $r \ll 1$. In this case, the   energy
density in an inertial frame is almost sinusoidal, with the positive energy outweighing the negative
energy only in order $r^2$. The effect of the periodic motion of the accelerated observer is to
introduce Doppler shift factors which enhance the negative energy compared to the positive energy.
The accelerated observer then sees the negative energy blueshifted and the positive energy redshifted.
In the cases of perpendicular motion treated in Sect.~\ref {sec:perpmotion} and \ref{sec:perpcav},
the effect is a transverse Doppler shift, and is hence of order $A^2$, where $A$ is the oscillation
amplitude.   For the parallel motion cases in Sects.~\ref{sec:parmotion} and \ref{sec:parcav}, the
effect is a linear Doppler shift, leading to an effect of order $A$. It is possible to have growing
negative integrated energy density even for arbitrarily slow motion, which means arbitrarily small 
$A$. However, for a given $A$, the squeeze state parameter $r$ is constrained by relations such as
Eqs.~(\ref{eq:Alim}) and (\ref{rlimpar}), which limit the rate of growth. Note that non-relativistic 
motion is not a requirement for growing negative energy, and the numerically integrated
results depicted in Figs.~\ref{fig:perpfig1pw} and \ref{fig:oscpwFig3} are for relativistic motion, but 
small squeeze parameter.

We studied a model which gives an operational meaning to integrated negative energy density
in the form of defocussing of bundle of worldlines. In Sect.~\ref{sec:focus}, we analyzed the
Raychaudhuri equation for the expansion along a bundle of accelerated worldlines. The
motivation for this study is that positive energy leads to attractive gravitational effects
and hence focussing, so negative energy should do the opposite. This expectation was born
out in our results. However, the situation is complicated by the need to include an
acceleration term in the Raychaudhuri equation, and the effects of the fluctuating velocity
of the accelerating charged particles in a fluctuating electromagnetic field. In the cases
which we examined, the gravitational effects and the acceleration effects have the same 
functional form.

The effect treated in this paper bears a superficial resemblance to the effect treated in
Ref.~\cite{PF11}, which is a linearly growing or decreasing mean squared velocity of a
charged particle undergoing sinusoidal motion near a mirror. The latter effect can be
interpreted as non-cancellation of anti-correlated quantum electric field fluctuations.
A charge at rest in the Casimir vacuum produced by the mirror is subjected to field
fluctuations which can give or take energy from the charge for a time consistent with
the energy-time uncertainty principle, but this effect will be cancelled by a subsequent
anti-correlated fluctuation. The sinusoidal motion upsets this cancellation, and allows
the mean squared velocity to grow or decrease, depending upon the phase of
the oscillation. The effect discussed in the present paper also involves linear growth, 
but does not have an obvious interpretation in terms of non-cancelling fluctuations.
The natural interpretation seems to be in terms of  Doppler shifts which can be
arranged to enhance negative energy and suppress positive energy.  A topic for 
future research is to study further the connection between these two effects.

Another topic is to understand to relation between the growing integrated negative
energy and the general worldline quantum inequality of Fewster~\cite{Fewster00}.
This inequality is difficult to evaluate explicitly for the sinusoidal worldline considered
here. In this case, the inequality must be weak enough to allow the linear growth
found here, but it might provide insight into the allowed behavior in situations
more general than we have treated.
 
A further question of interest is the possible physical consequences of 
accumulating
negative energy beyond those discussed in Sect.~\ref{sec:focus}. A possible
detection model for negative energy was proposed in Ref.~\cite{FR09}, in
which negative energy can suppress the decay rate of atoms in excited
states. The atoms in this model are moving along inertial worldlines, but
it might be possible to devise a more general model involving non-inertial motion.

Let us also stress that our results  {\em do not  in any way invalidate 
or diminish the 
implications of the quantum inequality bounds for inertial observers}. 
The strength of a quantum inequality bound may depend on the particular 
observer chosen, but the validity of the bound does not. As an example, 
suppose one is using a quantum inequality, applied to the motion of a 
particular inertial observer, 
to determine constraints on the geometry of a traversable wormhole. Let 
us further assume that in this case, the quantum inequality provides a very 
strong constraint. Now suppose one looks at the same problem from the point of 
view of, say, a different inertial or an accelerating observer and finds a much 
weaker bound. The weakness of the latter bound does not invalidate the strength 
of the previous bound. The observer whose motion provides 
the strongest quantum inequality bound implies the strongest constraint 
on the geometry of the wormhole. The latter cases simply yield true but 
weaker bounds.

 \begin{acknowledgments}
One of us (TR) would like to thank Werner Israel for a discussion, 
many years ago, of the moving mirror problem. The authors would also like 
to thank participants in the Beyond conference in January 2013  
for stimulating comments.
This work was supported in part by the National
Science Foundation under Grants PHY-0855360 and PHY-0968805.
\end{acknowledgments}

\end{document}